\documentclass{accordpaper}
\usepackage{amsmath}
\usepackage{pgfplots}
\pgfplotsset{compat=1.18}

\title{Accord: Schelling-Point Arbitration as a Composable Solana Primitive}
\author{Dr.~Fabian Schuh%
\thanks{Accord · useaccord.xyz · github.com/xeroc/accord. Manuscript
October 2026.}}
\date{2026-10}

\begin{document}

\maketitle

\begin{abstract}
Accord is an arbitration oracle on Solana. Any Solana program can create an
application specific court (i.e. subaccord) and file disputes. A random jury of
a specialized, stake-weighted juror pool reviews encrypted evidence and votes
under commit--reveal. Jurors that vote with the final ruling earn
fees and slashed stake. Jurors that vote against it lose stake. Appeals
grow the panel stepwise from 3 to 31 jurors, which raises both the cost
of maintaining a captured verdict and, under certain conditions, the
probability of a correct majority. Each pool defines its own collateral mint
and its own fee mint, so the economics can fit the application. This paper
elaborates the mechanism, the incentive model, the residual trust
assumptions, and the differences to the established court on EVM, Kleros. We
further discuss implementation details that are Solana specific.
\end{abstract}

\begin{IEEEkeywords}
arbitration, Schelling point, Solana, oracle, commit--reveal, sortition,
decentralized justice, dispute resolution
\end{IEEEkeywords}

\section{Introduction}
\IEEEPARstart{S}{mart} contracts execute agreed code and do not judge. No
program can decide by itself whether a freelancer met a milestone, whether a
token mint is for a genuine project, or whether a finding of an exploit was in
scope of a bounty. Further, we have seen multiple occurrences on-chain where
disagreement was handled poorly because no mechanism existed to address a
dispute properly. Examples are SushiSwap's "Chef Nomi" Incident (September
2020), roughly \$13–14 million worth of ETH from the project's developer fund
were unilaterally liquidated, the DAO hack in 2016 that caused Ethereum to fork
hard, but also oracle manipulation in 2025 where members of UMA could not agree
on a suit. These questions often carry value and not only need a binding answer
but also an established and trustworthy approach of reaching a verdict.

Today, Solana applications answer such questions with a trusted committee or
with no answer at all. A multisig can approve, deny, or drain. A recent wave of
treasury losses came from compromised signers that approved malicious payloads.
What teams want is a middle path: keep the authority, but let an independent
verdict control each privileged action.

Kleros gave Ethereum a working answer \cite{kleros-short,kleros-long}. It draws
random jurors, pays them for coherence with the majority, and slashes them for
deviation. It has settled more than 1{,}000 disputes since 2019. Solana
\cite{solana} has no native equivalent.

This paper presents Accord, a Solana-native Schelling-point arbitration oracle.
The design follows Kleros in its economics and departs from it in the draw, the
token model, the integration surface, and the failure
handling (Section~\ref{sec:deviations}). The paper contributes:
\begin{itemize}
  \item an on-chain stake accumulator with stake-weighted, verifiable,
    distinct-juror sortition driven by an oracle-authenticated VRF
    (Section~\ref{sec:draw});
  \item a two-mint, per-pool economic model that separates collateral
    from compensation (Section~\ref{sec:econ});
  \item a party-agnostic integration surface of two CPI instructions
    (Section~\ref{sec:interface});
  \item a stated trust profile with a security-value ceiling
    (Section~\ref{sec:model});
  \item reference applications built on the primitive
    (Section~\ref{sec:apps}).
\end{itemize}

Accord is an arbitration oracle. It emits a ruling. Enforcement stays
with the integrating application.

\section{Background and related work}

\subsection{Schelling points and the SchellingCoin}
Schelling described focal points: solutions that people choose without
communication because each person expects the other to choose them
\cite{schelling}. Buterin applied this to oracles as the SchellingCoin
\cite{schellingcoin}. Every token holder votes on a factual question by
secret ballot. After the reveal, the protocol rewards the majority and
penalizes the minority. For a holder $i$ with balance $h_i$ and vote
$v_i$, the basic game pays
\begin{equation}
  \pi_i \;=\;
  \begin{cases}
    +\,\delta\, h_i, & v_i = \mathrm{mode}(\mathbf{v}), \\[2pt]
    -\,\delta\, h_i, & v_i \ne \mathrm{mode}(\mathbf{v}),
  \end{cases}
  \label{eq:schelling}
\end{equation}
with redistribution rate $\delta$ and $\mathrm{mode}()$ denoting how to get to
the coherent outcome from the vector of all votes $\mathbf{v}$ . Each voter
maximizes its payoff by voting the answer it expects from the others. For
objective questions this recursion terminates at the true answer. Truth becomes
the focal point. Prior systems applied the same principle to oracles and
prediction markets \cite{truthcoin,augur}.

\subsection{Kleros}
Kleros is a decision protocol built on this mechanism
\cite{kleros-short,kleros-long}. Users stake a protocol token (PNK) into
a tree of subcourts. A filer submits a dispute to one subcourt. The
protocol draws jurors with probability proportional to stake, collects
commit--reveal votes, and rules by plurality. Coherent jurors earn
arbitration fees plus the stakes of incoherent jurors. Each appeal
doubles the panel plus one, which grows the cost of bribery
exponentially. Kleros adds a fork mechanism as a last defense against a
51\,\% token majority, and token-vote governance for policies and
parameters. As we will see below, Accord is resistant to these kind of
token-related attacks.

\subsection{Oracles with a vote fallback}
Optimistic oracles, such as UMA's data verification mechanism, propose
an answer and only vote on a challenge \cite{uma}. The vote is
token-weighted and has no drawn jury. In March 2025, one actor that
held about 25\,\% of UMA vote power changed the outcome of a
\$7\,million Polymarket market \cite{uma-incident,uma-orochi}. Weighted
plutocracy is a different mechanism from a drawn, slashed, appealable
jury. ERC-792 standardized the arbitrable/arbitrator split on Ethereum
\cite{erc792}. Accord carries the same split to Solana as a CPI
interface. It further only uses the stake-weight for drawing jurors. Once a
juror enters a jury, they all have equal say in the verdict.

\subsection{Accord deviations from Kleros}
\label{sec:deviations}
Accord inherits the incentive core of Kleros and changes not only the
tokenomics, but also those parts that a Solana implementation needs changed.
Further, Accord is a protocol to establish custom courts, e.g. subaccords to
serve specific needs and does not provide a fixed court tree.
Table~\ref{tab:deviations} lists the deltas in more detail.

\begin{table*}[!t]
\caption{Kleros versus Accord.}
\label{tab:deviations}
\centering
\small
\begin{tabular}{@{}p{68mm}p{90mm}@{}}
\toprule
\textcolor{ink}{\ttfamily\scriptsize KLEROS} &
\textcolor{ink}{\ttfamily\scriptsize ACCORD} \\
\midrule
\multicolumn{2}{@{}l}{\textcolor{ink}{\ttfamily\scriptsize Platform}} \\
Ethereum L1, L2 planned \cite{kleros-long} & Solana BPF program \\
\addlinespace
\multicolumn{2}{@{}l}{\textcolor{ink}{\ttfamily\scriptsize Court model}} \\
tree of subcourts, one General Court root &
flat Subaccords, permissionless creation, no root court \\
\addlinespace
\multicolumn{2}{@{}l}{\textcolor{ink}{\ttfamily\scriptsize Token}} \\
PNK, one token for stake and fees &
two arbitrary mints per Subaccord: staking token (collateral) and
fee token (compensation) \\
\addlinespace
\multicolumn{2}{@{}l}{\textcolor{ink}{\ttfamily\scriptsize Randomness}} \\
blockhash, VDF planned \cite{kleros-long} &
MagicBlock VRF \cite{magicblock}, oracle-authenticated
callback, one-shot and immutable per dispute \\
\addlinespace
\multicolumn{2}{@{}l}{\textcolor{ink}{\ttfamily\scriptsize Draw unit}} \\
PNK token; one juror can hold several votes (weight) &
juror; one seat per juror per round, distinct by construction \\
\addlinespace
\multicolumn{2}{@{}l}{\textcolor{ink}{\ttfamily\scriptsize Stake data}} \\
per-court token balances &
on-chain Merkle sum-tree accumulator, canonical root \\
\addlinespace
\multicolumn{2}{@{}l}{\textcolor{ink}{\ttfamily\scriptsize Votes}} \\
plurality, commit--reveal &
plurality or median on a scalar, commit--reveal with a
juror-bound preimage \\
\addlinespace
\multicolumn{2}{@{}l}{\textcolor{ink}{\ttfamily\scriptsize Appeals}} \\
panel $2N{+}1$, fee grows per round &
panel $2N{+}1$ up to 31, appeal bond plus a flip bounty \\
\addlinespace
\multicolumn{2}{@{}l}{\textcolor{ink}{\ttfamily\scriptsize Stake-majority defense}} \\
fork of the token \cite{kleros-long} &
none; the per-Subaccord staking mint limits the blast
radius of one capture \\
\addlinespace
\multicolumn{2}{@{}l}{\textcolor{ink}{\ttfamily\scriptsize Governance}} \\
PNK voting, liquid delegation planned &
per-Subaccord authority with a 48\,h timelock; no
protocol token \\
\addlinespace
\multicolumn{2}{@{}l}{\textcolor{ink}{\ttfamily\scriptsize Integration}} \\
ERC-792 arbitrable contracts \cite{erc792} &
two CPI instructions; wallets file disputes directly \\
\addlinespace
\multicolumn{2}{@{}l}{\textcolor{ink}{\ttfamily\scriptsize Juror gate}} \\
stake, Proof-of-Humanity mode planned \cite{kleros-long} &
stake; optional on-chain attestation gate per Subaccord \\
\bottomrule
\end{tabular}
\end{table*}

\section{System model and assumptions}
\label{sec:model}

\subsection{Roles}
The protocol has the on-chain roles and the off-chain roles of
Table~\ref{tab:roles}. Each role holds a specific power.

\begin{table*}[!ht]
\caption{Roles, powers, and residual assumptions.}
\label{tab:roles}
\centering
\small
\begin{tabular}{@{}p{68mm}p{90mm}@{}}
\toprule
\textcolor{ink}{\ttfamily\scriptsize FAILURE MODE} &
\textcolor{ink}{\ttfamily\scriptsize RESIDUAL ASSUMPTION} \\
\midrule
\multicolumn{2}{@{}l}{\textcolor{ink}{\ttfamily\scriptsize Filer}
  --- dispute content, options, evidence hash} \\
junk or wrong question & the filer states the question well \\
\addlinespace
\multicolumn{2}{@{}l}{\textcolor{ink}{\ttfamily\scriptsize Juror}
  --- one vote per seat} \\
lazy, incompetent, or bribed vote & honest majority of stake \\
\addlinespace
\multicolumn{2}{@{}l}{\textcolor{ink}{\ttfamily\scriptsize
  Credential issuer (optional)} --- pool admission via SAS gate} \\
junk or censored issuance & admission-only; issuer policy bounds
claims per human \\
\addlinespace
\multicolumn{2}{@{}l}{\textcolor{ink}{\ttfamily\scriptsize
  Subaccord authority} --- pool parameters via timelocked update} \\
hostile parameter change & a watcher reacts within 48\,h \\
\addlinespace
\multicolumn{2}{@{}l}{\textcolor{ink}{\ttfamily\scriptsize
  Evidence operator} --- confidential evidence delivery} \\
leak, selective service, forgery & operator does not leak or censor \\
\addlinespace
\multicolumn{2}{@{}l}{\textcolor{ink}{\ttfamily\scriptsize VRF provider}
  --- randomness for the draw} \\
withheld or delayed randomness & availability only; a stall forces a
refund, not a wrong ruling \\
\addlinespace
\multicolumn{2}{@{}l}{\textcolor{ink}{\ttfamily\scriptsize Indexer}
  --- Merkle paths for stake and draw} \\
withheld paths & at least one honest indexer serves the pool \\
\addlinespace
\multicolumn{2}{@{}l}{\textcolor{ink}{\ttfamily\scriptsize Cranker}
  --- progress of draws, windows, settlement} \\
stall & anyone can crank; someone does \\
\addlinespace
\multicolumn{2}{@{}l}{\textcolor{ink}{\ttfamily\scriptsize
  Upgrade authority} --- program code, pause} \\
malicious upgrade & multisig quorum until the post-audit freeze \\
\addlinespace
\multicolumn{2}{@{}l}{\textcolor{ink}{\ttfamily\scriptsize
  Integrating app} --- enforcement of the ruling} \\
ignored ruling & out of protocol scope by design \\
\bottomrule
\end{tabular}
\end{table*}

\subsection{Threat model}
\label{sec:threat}
The adversary is economically powerful and cryptographically bounded.
It may create arbitrarily many pseudonymous juror keys, acquire
arbitrary amounts of the staking mint subject to market availability,
sit on panels, collude, bribe privately, appeal whenever it can pay
the tender, submit or withhold permissionless crank transactions,
observe all public state including the stake distribution before
filing, know the full panel after the draw, and, if it is a dispute
party, submit misleading evidence. It may also try to induce non-reveal
and push disputes to failure. The security claim is economic and
cryptographic, not behavioral: nothing assumes the attacker behaves.

Under the cryptographic assumptions of the implementation, the
attacker cannot forge a VRF callback for a committed seed, alter the
frozen randomness $\xi$, the frozen root $R_D$, or the frozen dispute
terms, seat a juror whose proof fails against $R_D$, give one juror
two seats in a round, alter or contradict an accepted commitment, or
convert an availability failure of the VRF, the indexer, or evidence
delivery into a wrong ruling.

There are two boundaries we want to address explicitly. First, the evidence
operator shapes what jurors see, but only for evidence the filer routes
through it: the dispute binds the evidence hash not the evidence itself,
and the filer may deliver through any other channel
(Section~\ref{sec:evidence}). Within the operator channel, protocol
integrity survives an operator attack, information-layer integrity does
not. Second, vote secrecy binds observers, not the juror itself:
commit--reveal hides committed votes and blocks copying, but a juror
that discloses its vote and salt issues a verifiable receipt of the
committed vote (Section~\ref{sec:attack}).

\subsection{Juror information quality}
\label{sec:accuracy}
Honesty and competence are different properties. An honest juror votes
its belief instead of taking a bribe; competence is the probability
that the belief identifies the objectively correct answer. For a
binary question, let
\begin{equation}
  p \;=\; \Pr\!\left[\text{honest juror selects the correct answer}\right]
  \label{eq:accuracy}
\end{equation}
and assume for analysis that honest jurors err independently with
common accuracy $p$. The probability that an odd panel of $N$ reaches
the correct majority is then
\begin{equation}
  P_{\mathrm{correct}}(N,p) \;=\;
  \sum_{k=(N+1)/2}^{N} \binom{N}{k}\, p^{k} (1-p)^{N-k},
  \label{eq:condorcet}
\end{equation}
the Condorcet jury expression \cite{condorcet}. If $p > 1/2$, growing
$N$ improves the panel; if $p = 1/2$ the panel has no information
advantage over a coin flip; if $p < 1/2$ larger panels amplify the error.
The appeal ladder of Section~\ref{sec:appeals} therefore does double
duty: it multiplies the economic cost of capture and, under the same
independence assumption, raises the probability that the final panel
is correct. Independence is an analytical approximation, not a
guarantee: shared information sources, common misinformation, evidence
shaped by one operator, and professional-group bias correlate errors
and shrink the effective panel. The credential gate of
Section~\ref{sec:admission} is the protocol's lever on $p$. Obviously, no
mechanism whatsoever creates competence.

\subsection{Residual assumptions}
The Schelling honesty equilibrium is conditional. It holds under an
honest majority of staked collateral---where honest means unbribed and
voting one's belief, not infallible; whether the belief identifies the
correct answer is the separate question of Section~\ref{sec:accuracy}.
Juror admission is key-level
pseudonymous \cite{sybil}; one human can hold several keys and several
stakes. Vote secrecy holds until reveal, against observers: without the
salt a commitment reveals nothing, and the juror key inside its
preimage stops one juror from copying another's, so on-chain
observation neither reads votes nor enables free riding. Secrecy does
not bind the juror itself: the commitment is public, so a juror that
discloses its vote and salt produces a verifiable receipt of the
committed vote at any time after the commitment lands. What stays
unprovable is a vote never committed, or another juror's vote
(Section~\ref{sec:attack}).

Randomness and evidence delivery are external services with the failure modes of
Table~\ref{tab:roles}. The protocol contains each availability failure
with a refund path: no withholding failure produces a wrong ruling.
Subverting the information layer itself---a misleading question, forged
evidence---is not an availability failure and stays inside the
residual trust of Table~\ref{tab:roles}.

\subsection{Security-value ceiling}
\label{sec:ceiling}
Capture of a dispute is priced, not impossible. An attacker controls an outcome
through a stake majority or through bribery. The cheaper path sets the
ceiling,
\begin{equation}
  C^{*} \;=\; \min\bigl(C^{\mathrm{stake}},\; C^{\mathrm{bribe}}\bigr).
  \label{eq:ceiling}
\end{equation}
Section~\ref{sec:attack} derives both terms. Applications must file
disputes only for values below $C^{*}$.

\section{An exemplary dispute}
\label{sec:example}
This section walks one dispute end to end. The parameters are examples and real
Subaccords set their own according to their needs.

A freelancing escrow holds 2{,}000 USDC. The client claims the
milestone was not delivered. The developer disagrees. The escrow
program files a dispute into the \id{webdev} Subaccord: options
\id{[release, refund]}, evidence hash $h_E$, and $J=3$ jurors at
$f=5$ USDC per juror. The filing tender is
$(J{+}1)f = 20$ USDC. As we will see later, one appeal-bounty unit $f$
is withheld to reward a vindicated appellant.

\begin{enumerate}
  \item The filing stores the options, freezes the pool terms onto the
    dispute, and banks 15 USDC into the round-0 fee pot and 5 USDC into
    the bounty pool.
  \item A permissionless request draws randomness from the VRF oracle.
    The oracle answer writes the random value and freezes the juror-set
    Merkle root.
  \item Three seat-landing transactions result in three distinct jurors,
    Each drawn juror stakes at least $S_{\min}=1{,}000$ units of the
    pool collateral which could be USDC, or any other token.
  \item The evidence operator decrypts the evidence and re-encrypts it
    per juror. Each juror verifies the clear text against $h_E$.
  \item For vote $v_j$, each juror commits
    $c_j = H(v_j \,\Vert\, \mathrm{salt}_j \,\Vert\, k_j)$.
  \item Once every commitment is in, or the commitment period has passed, each
    juror reveals $(v_j,\mathrm{salt}_j)$. The reveal window
    closes early when all three revealed.
    result. If no party appeals within the 72\,h window, the dispute
    becomes final and every round settles against the final ruling:
    each coherent juror
    earns $\phi = \lfloor 15/2 \rfloor = 7$ USDC plus a share of the
    slash. The incoherent juror loses
    $\lambda = \alpha S_{\min} = 200$ collateral units at
    $\alpha = 0.2$.
  \item The escrow program reads the ruling, sees \id{refund}, and
    pays 2{,}000 USDC back to the client. The filer reclaims the unused
    5 USDC appeal bounty.
\end{enumerate}

Total delay: about four days. Total cost to the filer: 15 USDC plus
transaction fees.

\section{The Accord mechanism}
\label{sec:mechanism}

\subsection{The Arbitrable interface}
\label{sec:interface}
A Solana program becomes an Arbitrable with two calls:

\begin{acccode}
create\_dispute(s, opts, evid\_hash, fee);

Option<u64> = accord.get\_ruling(d);
\end{acccode}

The filer is a plain signer. A program files through CPI and enforces
the ruling on read-back: release an escrow, flip a registry item, gate
an authority. A wallet files directly and reads the ruling from
on-chain state. Accord holds no position on the parties and no
knowledge of the question. All party economics live in the Arbitrable.

\subsection{Components}
\label{sec:components}
Table~\ref{tab:components} lists the components. On-chain state lives
in program-derived accounts. Each Subaccord owns two token vaults: a
stake vault for collateral and a fee vault for fees, bonds, and
bounties.

\begin{table*}[!t]
\caption{Components.}
\label{tab:components}
\centering
\small
\begin{tabular}{@{}p{58mm}p{100mm}@{}}
\toprule
\textcolor{ink}{\ttfamily\scriptsize DUTY} &
\textcolor{ink}{\ttfamily\scriptsize KEY DATA} \\
\midrule
\multicolumn{2}{@{}l}{\textcolor{ink}{\ttfamily\scriptsize Subaccord}} \\
juror pool: terms, economics, gates &
staking mint, fee mint, $S_{\min}$, $\alpha$,
windows, $\theta$, $\varepsilon$, aggregation rule,
authority, evidence operator, optional SAS credential
schema, accumulator root \\
\addlinespace
\multicolumn{2}{@{}l}{\textcolor{ink}{\ttfamily\scriptsize JurorStake}} \\
one juror position in one pool &
staked amount, pending slash and reward deltas, fee
credit, live-seat counter \\
\addlinespace
\multicolumn{2}{@{}l}{\textcolor{ink}{\ttfamily\scriptsize Dispute}} \\
one filed question and its frozen terms &
options or scalar mode, per-round evidence hashes,
state, round counter, fee ledger, bounty pool, frozen
root, VRF value \\
\addlinespace
\multicolumn{2}{@{}l}{\textcolor{ink}{\ttfamily\scriptsize Round}} \\
one panel of one round &
seat list, commitments, revealed votes, window
deadlines, tally, draw-attempt counter \\
\addlinespace
\multicolumn{2}{@{}l}{\textcolor{ink}{\ttfamily\scriptsize AppealBond}} \\
one appeal deposit &
appellant, fee-plus-bond amount, prior result, flip
reward \\
\addlinespace
\multicolumn{2}{@{}l}{\textcolor{ink}{\ttfamily\scriptsize Stake vault}} \\
collateral custody & SPL token account of the staking mint \\
\addlinespace
\multicolumn{2}{@{}l}{\textcolor{ink}{\ttfamily\scriptsize Fee vault}} \\
compensation custody & SPL token account of the fee mint \\
\addlinespace
\multicolumn{2}{@{}l}{\textcolor{ink}{\ttfamily\scriptsize
  Evidence daemon (off-chain)}} \\
decrypt and re-encrypt evidence per
juror; public registry of pool policy documents &
encrypted blobs, delivery keys \\
\addlinespace
\multicolumn{2}{@{}l}{\textcolor{ink}{\ttfamily\scriptsize
  VRF oracle (off-chain)}} \\
randomness for the draw & scoped per-program identity \\
\bottomrule
\end{tabular}
\end{table*}

\subsection{Dispute lifecycle}
\label{sec:lifecycle}
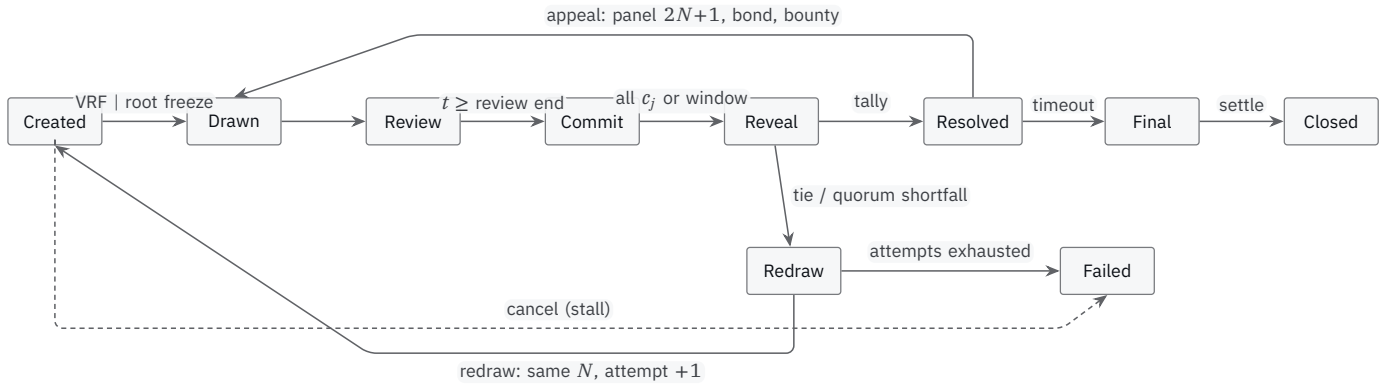
\begin{figure*}[!ht]
\centering
\resizebox{\textwidth}{!}{%

\begin{tikzpicture}[
  x=1.55cm, y=1.15cm,
  st/.style={
    draw=ink!55, line width=0.5pt,
    fill=paperlight,
    rounded corners=1.2pt,
    font=\sffamily\scriptsize,
    inner xsep=5pt, inner ysep=3.5pt,
    text=ink,
    minimum height=6.5mm,
    minimum width=13mm,
    align=center
  },
  tr/.style={
    -{Stealth[length=2.0mm,width=1.6mm]},
    draw=ink!65, line width=0.6pt,
    rounded corners=3pt
  },
  trd/.style={tr, dashed, dash pattern=on 2pt off 1.6pt},
  lb/.style={
    font=\sffamily\scriptsize,
    text=ink!80,
    inner sep=1pt,
    align=center,
    fill=paperlight,
    fill opacity=0.9,
    text opacity=1
  }
]

\node[st] (created)  at (0,0)    {Created};
\node[st] (drawn)    at (1.6,0)  {Drawn};
\node[st] (review)   at (3.2,0)  {Review};
\node[st] (commit)   at (4.8,0)  {Commit};
\node[st] (reveal)   at (6.4,0)  {Reveal};
\node[st] (resolved) at (8.2,0)  {Resolved};
\node[st] (final)    at (9.8,0)  {Final};
\node[st] (closed)   at (11.4,0) {Closed};

\node[st] (redraw) at (6.6,-1.8) {Redraw};
\node[st] (failed) at (9.4,-1.8) {Failed};

\draw[tr] (created)  -- node[lb,above=1mm]{VRF $\vert$ root freeze} (drawn);
\draw[tr] (drawn)    -- (review);
\draw[tr] (review)   -- node[lb,above=1mm]{$t \ge$ review end}     (commit);
\draw[tr] (commit)   -- node[lb,above=1mm]{all $c_j$ or window}    (reveal);
\draw[tr] (reveal)   -- node[lb,above=1mm]{tally}                  (resolved);
\draw[tr] (resolved) -- node[lb,above=1mm]{timeout}                (final);
\draw[tr] (final)    -- node[lb,above=1mm]{settle}                 (closed);

\draw[tr] (reveal) -- node[lb,right=1mm]{tie / quorum shortfall} (redraw);

\draw[tr] (redraw) -- node[lb,above=1mm]{attempts exhausted} (failed);

\draw[tr]
  (resolved.north) -- ++(0,0.75)
  -- node[lb,above=1mm,pos=0.5]{appeal: panel $2N{+}1$, bond, bounty}
  ++(-5.0,0) -- (drawn.north);

\draw[tr]
  (redraw.south) -- ++(0,-0.7)
  -- node[lb,below=1mm,pos=0.5]{redraw: same $N$, attempt $+1$}
  ++(-3.8,0) -- (created.south);

\draw[trd]
  (created.south) -| ++(0,-2.2)
  -- node[lb,above=1mm,pos=0.5]{cancel (stall)}
  ++(9,0) -- (failed.south);

\end{tikzpicture}%

}
\caption{Dispute state machine. All rounds settle at finality
against the final ruling. A pre-draw VRF stall, a post-draw
stall past its grace period, or exhausted redraw attempts move the
dispute to Failed. Fees and bonds refund, and no ruling exists.}
\label{fig:statemachine}
\end{figure*}

Figure~\ref{fig:statemachine} shows the dispute state machine.
Table~\ref{tab:instructions} lists the lifecycle operations. Every
state advance after a filing is a permissionless crank: anyone can move
a dispute forward when its window elapses. Several windows close early
when the panel is complete.

\begin{table*}[!ht]
\caption{Dispute-lifecycle operations. Secondary operations (fee
withdrawal, bond and bounty claims, stall cancellation, pool
maintenance, the circuit breaker) are permissionless cranks and are
omitted.}
\label{tab:instructions}
\centering
\small
\begin{tabular}{@{}p{44mm}p{26mm}p{92mm}@{}}
\toprule
\textcolor{ink}{\ttfamily\scriptsize OPERATION} &
\textcolor{ink}{\ttfamily\scriptsize ACTOR} &
\textcolor{ink}{\ttfamily\scriptsize RESULT} \\
\midrule
Found a juror pool & anyone & new pool with terms, two mints, empty
accumulator \\
Join or leave the pool & juror & collateral moves in or out; selection
weight updates live \\
File a dispute & filer: program or wallet & question, options, evidence
hash recorded; terms frozen; tender paid \\
Draw randomness & anyone & oracle request bound to this dispute; the
oracle delivers the value and freezes the pool snapshot \\
Land a seat & anyone & one juror seated; membership and sortition
verified against the frozen root \\
Commit a vote & drawn juror & hash commitment recorded \\
Reveal a vote & drawn juror & vote and salt opened and recorded \\
Tally a round & anyone & result recorded under quorum and decisiveness
gates, or the round marked for redraw \\
Appeal & anyone & new round at panel $2N{+}1$; fee, bond, and bounty
paid \\
Finalize and settle & anyone & final ruling fixed; every round settled
against it \\
Redraw the panel & anyone & same-size panel re-seated; no-shows
slashed \\
Read the ruling & anyone & final ruling, or none before finality \\
\bottomrule
\end{tabular}
\end{table*}

\subsection{Juror admission}
\label{sec:admission}

Stake admits keys, not qualifications. Juror admission is key-level
pseudonymous (Section~\ref{sec:limitations}), so a pool that wants
competence---or a bound on how cheaply one human can hold many
seats---needs a filter that collateral does not provide. A Subaccord
can therefore bind an optional admission gate at creation: a credential
schema of the Solana Attestation Service (SAS) plus an issuer allowlist
\cite{solana-attestation}. Like the mints and the aggregation rule, the
gate is immutable from creation.

SAS anchors verifiable claims on chain. An issuer---a professional
society, a proof-of-personhood provider, the integrating application
itself---publishes a schema and issues attestations that bind a claim
to a holder key, with an expiry, in a program-owned account. When the
gate is set, the join instruction verifies the juror's attestation
through a CPI to the SAS program: schema match, issuer on the
allowlist, holder key equal to the staking key, attestation unexpired.
The seat-landing call re-checks the same facts, so an expired or
revoked attestation keeps a juror out of new panels. Verification reads
public account state; no issuer involvement is needed at draw time.

The gate prices sybil admission, it does not eliminate it. Nothing on
chain stops one human from holding several keys; each key now needs its
own unexpired attestation, and the issuer's issuance policy---claims
per human, diligence behind each claim---becomes the bound.
Zero-knowledge presentations let a juror prove the schema's statement
without revealing the underlying identifier, so a gate can demand
``licensed dentist per issuer $D$'' or ``unique human per issuer $P$''
without the pool learning more. Sortition weight is untouched: the
accumulator still weighs stake only, and the gate decides who may hold
it. A coalition that wants $k$ seats now needs $k$ attestations instead
of $k$ keys, at issuer prices.

The residual assumption moves but does not shrink. A junk issuer admits
junk jurors; a censoring issuer bars honest ones; an issuer that goes
offline stops new admissions and lets the pool drain as attestations
expire. The gate is admission-only by construction: no ruling, fee, or
slash path depends on the issuer at dispute time, so issuer failure is
a liveness failure of the pool and not a wrong ruling---the same
containment the protocol applies to randomness and evidence.

\subsection{Jury selection}
\label{sec:draw}
\subsubsection{The stake accumulator}
\label{sec:accumulator}
Sortition must pick jurors at random, weight the pick by stake, and
prove the pick on-chain. The straight-forward design would load every juror
account into the draw transaction and reads the stakes directly. Solana forbids
this: a legacy transaction caps at 1{,}232 bytes and about 35 accounts
\cite{anza-vtx}, and lookup-table compression still caps the pool in the
hundreds. A pool worth attacking needs orders of magnitude more jurors than any
transaction can address.

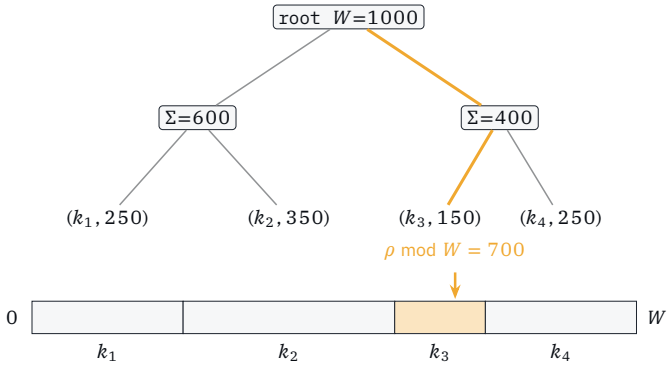
\begin{figure}[!ht]
\centering
\resizebox{\columnwidth}{!}{%
\begin{tikzpicture}[x=1mm, y=1mm,
  nd/.style={font=\ttfamily\scriptsize, text=ink, inner sep=1.6pt},
  sum/.style={draw=hairline, fill=paperlight, rounded corners=1.5pt,
              font=\ttfamily\scriptsize, text=ink, inner sep=2pt},
  edge/.style={draw=ink!45, line width=0.6pt},
  pth/.style={draw=amber, line width=1.2pt}]
\node[sum] (root) at (44,28) {root $W{=}1000$};
\node[sum] (p12) at (23,14) {$\Sigma{=}600$};
\node[sum] (p34) at (65,14) {$\Sigma{=}400$};
\node[nd] (k1) at (10.5,0)  {$(k_1,250)$};
\node[nd] (k2) at (35.7,0)  {$(k_2,350)$};
\node[nd] (k3) at (56.7,0)  {$(k_3,150)$};
\node[nd] (k4) at (73.5,0)  {$(k_4,250)$};
\draw[edge] (root) -- (p12);
\draw[pth]  (root) -- (p34);
\draw[edge] (p12) -- (k1);
\draw[edge] (p12) -- (k2);
\draw[pth]  (p34) -- (k3);
\draw[edge] (p34) -- (k4);
\draw[hairline, fill=paperlight] (0,-16) rectangle (21,-11.5);
\draw[hairline, fill=paperlight] (21,-16) rectangle (50.4,-11.5);
\draw[hairline, fill=amber!30] (50.4,-16) rectangle (63,-11.5);
\draw[hairline, fill=paperlight] (63,-16) rectangle (84,-11.5);
\node[nd] at (10.5,-18.6) {$k_1$};
\node[nd] at (35.7,-18.6) {$k_2$};
\node[nd] at (56.7,-18.6) {$k_3$};
\node[nd] at (73.5,-18.6) {$k_4$};
\draw[-{Stealth[length=2mm]}, amber, line width=1pt]
  (58.8,-7.5) -- (58.8,-11.2);
\node[font=\sffamily\scriptsize, text=amber] at (58.8,-4.6)
  {$\rho \bmod W = 700$};
\node[nd] at (-2.8,-13.75) {$0$};
\node[nd] at (86.8,-13.75) {$W$};
\end{tikzpicture}}%
\caption{The stake accumulator. Leaves partition $[0, W)$ into
stake-proportional intervals; a seat value lands in one interval. The
amber seat lands on $k_3$: its interval $[600, 750)$ is reconstructed
from the two sibling sums alone, and the sibling digests hash the path
back to the frozen root. Verification cost per seat is
$O(\log N)$, independent of pool size.}
\label{fig:mst}
\end{figure}

Accord therefore keeps the pool in a Merkle sum tree (Figure~\ref{fig:mst}). A
leaf holds a juror key and its live stake. An inner node holds its children's
hash and their stake sum. The root is canonical on-chain: every stake movement
updates one leaf through a caller-supplied path in $O(\log n)$ accounts and
recomputes the root. There is no snapshot, no posting bond, and no challenge
window required for this update.

The live accumulator and each draw's snapshot are separate objects.
At the VRF callback the dispute freezes its draw parameters in one
shot: the random value $\xi$, the juror-set root
$R_D = R_{\mathrm{pool}}(t_{\mathrm{freeze}})$, and the total stake
$W_D$ under that root. Every seat of the dispute proves membership
against exactly $(R_D, W_D)$, and not against the moving pool state.
Stake may still move after the freeze---the accumulator serves other
disputes---but movement cannot touch a frozen draw: a landing call
re-checks the juror's live stake against $S_{\min}$ and rejects the
seat if it has fallen below, sending the crank to a retry, and a
withdrawal that would drop a juror below the locked exposure of a live
seat is rejected outright.

The tree itself is canonical. Leaves are ordered by juror public key,
one live leaf per key; a zero-stake leaf carries zero selection weight
and cannot be landed, existing only as structure. Stake sums are
fixed-width integers with an explicit cap on $W$; an update that would
overflow the accumulator type reverts. The indexer that supplies
Merkle paths cannot forge membership: the program authenticates every
caller-supplied path against $R_D$, so a malicious indexer can withhold
service or waste fees, but cannot seat an unproven juror.

The leaves, read in tree order, partition $[0, W)$ into intervals proportional
to stake, and a seat value $\rho \bmod W$ lands in exactly one interval. To
verify the landed juror, the chain needs that leaf plus one sibling digest per
level: each sibling sum supplies the missing part of the stake prefix $P_j$ of
\eqref{eq:interval}, and the sibling digests hash the path back to the frozen
root. One seat costs one account plus $\lceil \log_2 N \rceil$ digests, whatever
the pool size. In Figure~\ref{fig:mst} the amber seat lands on $k_3$: the
sibling sums 600 and 250 reconstruct its interval $[600, 750)$ without reading
$k_1$, $k_2$, or $k_4$.

A sibling digest costs 40 bytes: 32 for the hash and 8 for the stake
sum. With packet budget $B$ and fixed overhead $F$ of about 300 bytes
(signature, header, blockhash, core accounts, call data), the tree
depth that fits one seat transaction is
\begin{equation}
  d \;\le\; \left\lfloor \frac{B - F}{40} \right\rfloor,
  \qquad
  \left\lvert \mathcal{S} \right\rvert \;\le\; 2^{d},
  \label{eq:capacity}
\end{equation}
which turns a hard platform limit into a pool-size guarantee.
Table~\ref{tab:capacity} evaluates it. A legacy packet carries depth
23, about $8.4 \times 10^{6}$ seats. The v1 transaction format raises
the packet to 4{,}096 bytes \cite{solana-v1} and carries depth 94,
about $2 \times 10^{28}$ seats. The protocol default depth is 20,
about one million seats, which leaves compute headroom inside a legacy
packet today. Under v1 the packet stops binding at all: rent economics
and per-level hashing become the limit, not the transaction format.

\begin{table}[!t]
\caption{Pool capacity per transaction format, from
\eqref{eq:capacity} with $F = 300$ bytes.}
\label{tab:capacity}
\centering
\small
\begin{tabular}{@{}p{20mm}p{18mm}p{12mm}p{22mm}@{}}
\toprule
\textcolor{ink}{\ttfamily\scriptsize FORMAT} &
\textcolor{ink}{\ttfamily\scriptsize PACKET} &
\textcolor{ink}{\ttfamily\scriptsize DEPTH} &
\textcolor{ink}{\ttfamily\scriptsize POOL SEATS} \\
\midrule
legacy (v0) & 1{,}232 B & 23 & $2^{23} \approx 8.4$\,M \\
v1          & 4{,}096 B & 94 & $2^{94} \approx 2 \times 10^{28}$ \\
protocol default & any  & 20 & $2^{20} \approx 1$\,M \\
\bottomrule
\end{tabular}
\end{table}

\subsubsection{Randomness}
A permissionless request calls the MagicBlock VRF program
\cite{magicblock} with the dispute key as the request seed and the
Accord callback as the only destination. The oracle answers through a
signature of a per-program identity PDA, a construction in the family
of verifiable random functions \cite{chainlink-vrf}. The callback
writes $\xi$ and freezes the juror-set root in one shot. A second
request or callback for the same dispute reverts. No caller input can
influence $\xi$ after the commit, so the brute-force attack of
re-rolling randomness until a favorable panel appears is closed at the
instruction level.
The provider is trusted for liveness, not for integrity: a VRF output
arrives with a proof that the chain verifies against the provider's
authenticated identity, and for a committed seed no different output
verifies without breaking the underlying signature scheme. Withholding
and delay are therefore the provider's only levers, and a timeout
entitles any caller to cancel the dispute and trigger a filer refund.

\subsubsection{Sortition}
One VRF value serves the whole dispute. The protocol derives one
random value per seat, salted by round, draw attempt, seat, and retry,
\begin{equation}
  \rho_{s} \;=\;
  H\!\left(\xi \,\Vert\, D \,\Vert\, i \,\Vert\, a \,\Vert\, s \,\Vert\, \tau\right),
  \label{eq:seed}
\end{equation}
with dispute $D$, round $i$, draw attempt $a$, seat $s$, and retry
$\tau$. Let $P_j$ be the stake prefix of juror $j$ in tree order and
$W$ the total stake. Seat $s$ lands on juror $j^{*}(s)$ when
\begin{equation}
  P_j \;\le\; \rho_{s} \bmod W \;<\; P_j + w_j .
  \label{eq:interval}
\end{equation}
Each seat therefore selects juror $j$ with probability
\begin{equation}
  \Pr\bigl(j^{*}(s) = j\bigr) \;=\; \frac{w_j}{W},
  \qquad
  \Pr\bigl(j \in \mathcal{J}_i\bigr)
  \;\approx\; 1 - \left(1 - \tfrac{w_j}{W}\right)^{N_i},
  \label{eq:sortition}
\end{equation}
where $\mathcal{J}_i$ is the panel of round $i$ and $N_i$ its size.
Equation~\eqref{eq:sortition} holds when seat draws do not collide.
A collision increments $\tau$, re-derives the seat value, and so
enforces a distinct panel: one juror holds at most one seat per round.
Kleros instead draws tokens and gives one juror several votes
\cite{kleros-long}. Accord draws jurors, which keeps one stake
position equal to one vote and makes every panel member individually
slashable.

Equation~\eqref{eq:sortition} is the independent-draw approximation.
The collision retry turns panel selection into weighted sampling
without replacement: inclusion probabilities run slightly above the
independent formula for heavy jurors, because a retried seat redraws
among the unseated only, and the capture analysis of
Section~\ref{sec:attack} inherits the same approximation. The offline
evaluation (Section~\ref{sec:evaluation}) runs the exact sampler and
bounds the deviation: single-digit percent for moderate stakes, up to
14 percent for the heaviest juror of a Zipf-weighted test pool.

\subsubsection{One seat per transaction}
A membership proof for one seat fills a large part of a packet, so one
transaction lands exactly one seat, and a crank repeats it $N_i$ times.
Each call verifies the Merkle membership proof against the frozen root
and the sortition interval of \eqref{eq:interval} against the
authenticated sibling sums. The call then increments the juror's
live-seat counter, which locks the stake in place for the round.

\subsection{Evidence}
\label{sec:evidence}
On-chain state stores hashes only. The filing commits the round-0 evidence
hash $h_E$, and each appeal can commit a new hash for its round. Once
committed, a hash is immutable: it binds the dispute to exactly one
evidence file set, and nothing else about the evidence---no blob, no
storage location, no delivery path---is protocol state.

Each Subaccord names an evidence-operator public key at creation. That key
defines the confidentiality channel: the filer encrypts the evidence file
set to the operator key and puts the encrypted blobs into any off-chain
storage. The operator is the only entity able to decrypt them; after the
draw it decrypts and re-encrypts the evidence to each drawn juror key. The
channel buys confidentiality and convenience---the filer encrypts once,
before the jury is known, and the plain text never appears in public
storage---at the price of the operator seeing it.

The channel is the filer's choice, not a protocol requirement. Because the
dispute references the hash only, the filer may instead publish the file
set, for example on IPFS, or wait for the draw and deliver the evidence
itself, encrypted to the drawn juror keys. A filer that delivers directly
removes the operator from the loop entirely.

Whatever the channel, the hash is the authenticity check: a juror accepts
evidence only if its clear text hashes to the committed $h_E$. No delivery
party, the operator included, can substitute content on a verifying juror;
a forged delivery fails the check and degrades to withholding. The
residual operator trust on routed evidence is leak and selective service,
as stated in Table~\ref{tab:roles}, and is the target of future work
(Section~\ref{sec:future}).

\subsection{Votes and aggregation}
\label{sec:votes}
Each drawn juror commits
\begin{equation}
  c_j \;=\; H\!\left(\mathrm{le8}(v_j) \,\Vert\, \mathrm{salt}_j
         \,\Vert\, k_j\right),
  \label{eq:commit}
\end{equation}
with the vote $v_j$ as an 8-byte little-endian integer and $k_j$ the
juror key. The key inside the preimage stops one juror from copying
the commitment of another. The reveal opens $(v_j,\mathrm{salt}_j)$.
The protocol verifies $c_j$ and records the vote. The reveal window
closes early when every juror revealed.

A Subaccord picks one of two aggregation rules, frozen onto the
dispute at filing.

\subsubsection{Plurality}

In this section, we specify the purpose of $\mathrm{mode}()$ as used earlier.
The dispute files 2 to 8 options. A vote is an option index. The round tally is
the mode,
\begin{equation}
  r_i \;=\; \operatorname*{arg\,max}_{o}\;
  n_o, \qquad
  n_o \;=\; \sum_{j \in \mathcal{J}_i}
  \mathbf{1}\!\left[v_j = o\right].
  \label{eq:plurality}
\end{equation}
A tie at the top count is a non-decisive round. The round redraws
rather than pick an arbitrary winner. Odd panels remove ties only for
binary votes with a full reveal. A split or partial panel can still
tie, and still redraws.

\subsubsection{Median}
The dispute files no options. A vote is an unsigned scalar, for
example a price or a percentage in settlement-mint base units. The
round tally is the upper median of the revealed values,
\begin{equation}
  r_i \;=\; V_{(\lfloor n/2 \rfloor)},
  \qquad V_{(0)} \le V_{(1)} \le \dots \le V_{(n-1)},
  \label{eq:median}
\end{equation}
which stays deterministic when non-reveals shrink the panel to an even
size. Scalar voting drops the clone problem of plurality: near answers
do not split the vote, they pull the median.

\subsubsection{Reveal quorum}
A round resolves only when enough jurors revealed,
\begin{equation}
  n_{\mathrm{rev}} \;\ge\;
  \left\lceil \theta\, N_i \right\rceil,
  \qquad \theta = \tfrac{2}{3} \;\text{(default)} .
  \label{eq:quorum}
\end{equation}
A shortfall, or a tie under \eqref{eq:plurality}, marks the round
redraw-eligible. The protocol slashes each no-show by $\lambda$,
releases the seats, and redraws the same-size panel with a fresh draw
attempt $a$. After the draw-attempt cap (default 3) the dispute fails.
Fees and bonds refund, and no ruling exists. A large holder can force
a Failed dispute by silence, but pays
$\lambda \cdot \mathrm{seats} \cdot \mathrm{attempts}$ for it and
cannot force a wrong ruling.

\subsection{Ruling and incentives}
\label{sec:incentives}
\subsubsection{Coherence}
Coherence is judged against the final ruling $r$, not the round
tally. For plurality it is exact match. For median it is a tolerance
band,
\begin{equation}
  \mathrm{coh}(j) \;=\;
  \begin{cases}
    \mathbf{1}\!\left[v_j = r\right], & \text{plurality},\\[2pt]
    \mathbf{1}\!\left[\,\lvert v_j - r \rvert
      \le r \cdot \varepsilon\,\right], & \text{median},
  \end{cases}
  \label{eq:coherence}
\end{equation}
with per-pool tolerance $\varepsilon$ (default $0.01$, one percent).
Every round of the dispute, superseded ones included,
settles against the same $r$. A juror that votes with an overturned
majority of an early round recovers nothing. A juror in a vindicated
minority takes the whole pot of its round.

\subsubsection{Fees and slashing}
Let $n_c$ be the coherent count and $n_{\bar c}$ the incoherent count
of a round. Let the round pot be $\Pi_i = N_i f$ plus any forfeited
bonds. The protocol pays each coherent juror
\begin{equation}
  \phi_i \;=\; \left\lfloor \frac{\Pi_i}{n_c} \right\rfloor,
  \qquad
  \psi \;=\; \left\lfloor \frac{\lambda\, n_{\bar c}}{n_c} \right\rfloor,
  \qquad
  \lambda \;=\; \alpha\, S_{\min},
  \label{eq:pay}
\end{equation}
where $\phi_i$ is the fee share, $\psi$ the stake-redistribution
share, and $\lambda$ the flat slash of one incoherent juror. Division
remainders stay in the protocol as surplus. A juror that revealed but
voted incoherent pays $\lambda$. A juror that fails to reveal by the
deadline pays the same $\lambda$, decisive round or shortfall alike:
silence never prices below an honest reveal, and the forfeited fee
share returns to the pot. The net payoff per round is
\begin{equation}
  \pi_j \;=\;
  \begin{cases}
    \phi_i + \psi, & \mathrm{coh}(j) = 1,\\[2pt]
    -\lambda,      & \text{revealed, } \mathrm{coh}(j) = 0,\\[2pt]
    -\lambda,      & \text{no reveal by the deadline}.
  \end{cases}
  \label{eq:cases}
\end{equation}
If no juror is coherent but some revealed, the pools fall back to the
revealers. If none revealed, the pot stays in the vault as protocol
revenue.

\subsubsection{Why truth is the focal point}
The argument has three layers.

Individually, voting is monotone in belief. Let $p_o$ be the
probability a juror assigns to option $o$ equaling the final ruling.
With uniform no-show slashing, silence is never the better row of
\eqref{eq:cases}, so from \eqref{eq:pay} the expected payoff of
revealing $o$ is
\begin{equation}
  \mathbb{E}\!\left[\pi_j \mid v_j = o\right]
  \;=\; p_o\,(\phi_i + \psi) \;-\; (1 - p_o)\,\lambda ,
  \label{eq:expected}
\end{equation}
which rises strictly in $p_o$: the best response is to vote the
expected final winner. The analysis treats $p_o$ as exogenous, a
price-taking juror that cannot swing its panel---exact for large
panels, an approximation for round 0. Beliefs are anchored one level
up: the final ruling belongs to the largest, most expensive panel, so
each juror votes the verdict it expects at the deepest appeal it
expects.

Coordinately, the vote is a Schelling game. If jurors believe the
others are independently better than random at recognizing the correct
answer---$p > 1/2$ in Section~\ref{sec:accuracy}---the correct answer
is the highest-probability focal outcome, and \eqref{eq:expected} makes
it every honest juror's best response
\cite{schellingcoin,kleros-long}.

Statistically, the appeal ladder aggregates. Under independent honest
errors with $p > 1/2$, \eqref{eq:condorcet} raises the probability of
a correct majority as the panel grows from 3 to 31, while
Section~\ref{sec:attack} raises the cost of holding a wrong verdict in
place. The failure modes are the mirror image: $p = 1/2$ gives no
information advantage, $p < 1/2$ amplifies error, correlated errors
shrink the effective panel, and a stake majority can buy the outcome
outright. Honest jurors can still lose a single case. Over a portfolio,
fees and redistribution must exceed losses, or the pool is mispriced.


\subsection{Appeals}
\label{sec:appeals}
Round 0 holds $J$ jurors, $J$ odd (default 3). Each appeal opens a new
round of twice the panel plus one,
\begin{equation}
  N_0 = J, \qquad N_{i+1} = 2 N_i + 1
  \;\;\Longleftrightarrow\;\;
  N_i = (J{+}1)\,2^{i} - 1,
  \label{eq:ladder}
\end{equation}
with the series capped at 31 jurors currently ($J{=}3$: 3, 7, 15, 31). The
appellant pays, in fee tokens,
\begin{equation}
  T^{\mathrm{appeal}}_i \;=\;
  \underbrace{N_{i+1}\,f}_{\text{new round fee pot}} \;+\;
  \underbrace{N_{i+1}\,f}_{\text{appeal bond}} \;+\;
  \underbrace{f}_{\text{bounty unit}},
  \label{eq:tender}
\end{equation}
and the filer pays the same shape at filing,
\begin{equation}
  T^{\mathrm{file}} \;=\;
  \underbrace{J\,f}_{\text{round-0 fee pot}} \;+\;
  \underbrace{f}_{\text{bounty unit}} \;=\; (J{+}1)\,f .
  \label{eq:filetender}
\end{equation}
The bond refunds when the dispute ends. The bounty units accumulate in
the dispute bounty pool
\begin{equation}
  \Lambda \;=\; (1 + m)\, f,
  \label{eq:bounty}
\end{equation}
with $m$ the appeal count. If the dispute ends with no appeal, the
filer reclaims $\Lambda$. If an appeal changes the result and its own
round's verdict is the final ruling, the protocol splits $\Lambda$ onto
the aligned flip bonds: a vindicated appellant recovers its bond plus a
bounty. The bounty is a subsidy, not a guarantee of profit. $\Lambda$
grows linearly in $m$ while the tender of \eqref{eq:tender} grows as
$2^{m}$: at $J{=}3$ and $m{=}3$ the accumulated bounty is $4f$ against
a final-round tender of $63f$. Early appeals can be bounty-positive;
the last rung is value-driven, and a party appeals when correcting the
ruling is worth more than the tender.

One boundary sits outside the mechanism: nothing forces a losing party
to appeal. An attacker that benefits from a wrong round-0 ruling can
offer the victim a side payment to let the appeal window lapse, priced
against the victim's own expected gain from appealing. The ladder
therefore raises the cost of maintaining capture against an
economically rational challenger; it does not manufacture one.

\subsection{Security properties}
\label{sec:attack}

\subsubsection{No double draw}
The seed of \eqref{eq:seed} mixes the retry counter $\tau$. When a
seat lands on a juror that already holds a seat in the round, the draw
call rejects the seat, and the crank retries with $\tau + 1$. The
live-seat counter double-checks the invariant on-chain. One
juror, one seat, one slash exposure per round.

\subsubsection{Vote secrecy and pre-revelation}
Votes stay hidden between commit and reveal---from everyone the juror
does not arm with the preimage. The juror key inside the commitment
preimage of \eqref{eq:commit} stops copy votes: a second juror cannot
submit an existing $c_j$ as its own. But $c_j$ is public, so a juror
that discloses $(v_j, \mathrm{salt}_j)$, privately or publicly, hands
out a cryptographically verifiable receipt of its committed vote:
anyone recomputes \eqref{eq:commit} and compares against the on-chain
commitment. A receipt exists only for the committed value; the juror
cannot prove a vote it never committed. Commit--reveal therefore stops
passive observation and vote copying. It does not stop willing
disclosure: bribery and coordination inside the commit window can
settle on verifiable receipts. Full receipt-freeness is a stronger
property that Accord v1 does not claim. Kleros studied a scheme that
lets anyone steal the stake of a pre-revealing juror \cite{antipre},
which works precisely because a disclosed preimage verifies against
the public commitment. Version 1 of Accord relies on the commit
window bound alone: the reveal instruction opens only once every
commitment is in or the commit window has passed.
Adopting a penalty of that shape is future work
(Section~\ref{sec:future}).

\subsubsection{Collusion and bribery}
A drawn panel is unknown before the draw, so collusion can only form
after it. Its terms can be verifiable---a co-conspirator proves the
committed vote by disclosing its vote and salt---but every proof needs
the juror's own cooperation, and no juror can read or copy another's
commitment. Bribery prices from \eqref{eq:pay}: a
bought majority of round $i$ needs
$\lfloor N_i/2 \rfloor + 1 = (N_i{+}1)/2$ jurors on an odd panel, each
paid at least its forgone $\phi_i + \psi$ plus the slash $\lambda$ it
accepts. Buying one round and holding a verdict are different
purchases: a dispute only reaches round $i{+}1$ while the losing party
keeps appealing, so under the assumption that the victim appeals every
adverse round, an attacker that starts bribing in round 0 must hold
every panel through $m$, at a total cost of at least
\begin{equation}
  C^{\mathrm{bribe}}(m) \;=\;
  \frac{(J{+}1)\left(2^{\,m+1} - 1\right)}{2}\,\bigl(f + \lambda\bigr).
  \label{eq:bribe}
\end{equation}
With $J{=}3$ and $m{=}3$, equation~\eqref{eq:bribe} gives
$30\,(f + \lambda)$: the attacker must buy $2{+}4{+}8{+}16 = 30$
juror-seats across the four panels. The $p{+}\epsilon$ attack
\cite{pepsilon} promises the bribe only on failure and so costs
nothing when it works. The defense is the same escalation, plus the
observation that a captured final round destroys the value of the
collateral mint the attacker must hold. The converse
strategy---paying the victim not to appeal---is priced in
Section~\ref{sec:appeals}.

\subsubsection{Stake-majority capture}
Let $q = W_A / W$ be the attacker's share of staked collateral. Draws
are stake-weighted, so the attacker holds a majority
\emph{advantage}, not a certainty. Under the independent-draw
approximation, the probability of controlling a strict majority of a
panel of size $N$ is
\begin{equation}
  P_{\mathrm{capture}}(q, N) \;=\;
  \sum_{k=\lceil (N{+}1)/2 \rceil}^{N}
  \binom{N}{k}\, q^{k} (1-q)^{N-k},
  \label{eq:capture}
\end{equation}
plotted in Figure~\ref{fig:capture}. Below $q = 1/2$, larger panels
suppress accidental capture toward zero; above it, they harden it
toward certainty; at $q = 1/2$ every panel is a coin toss. Appeals do
not repair a stake majority: \eqref{eq:capture} applies afresh at
panel 31. The cost is the market cost of the majority position,
\begin{equation}
  C^{\mathrm{stake}} \;\approx\; \tfrac{1}{2}\, W\, q,
  \label{eq:stakecost}
\end{equation}
with $u$ the unit price of the staking mint. The net cost can sit far
below \eqref{eq:stakecost}, because the attacker also collects its own
fees and slashes. A pool must therefore price $C^{*}$ of
\eqref{eq:ceiling} conservatively. Two structural mitigations bound the
damage. First, the staking mint is per pool: one capture does not
cross pools, and a captured pool's collateral collapses in price while
the attacker still holds it. Second, any watcher can read stake
concentration from public juror positions and price the pool before
filing. Kleros answers the same attack with a token fork
\cite{kleros-long}. Accord accepts the smaller scope in exchange for
the simpler protocol.

\subsubsection{Veto by abstention}
A coalition larger than $1 - \theta$ can withhold reveals, force
the redraw ladder, and push the dispute to Failed. The outcome is a
refund, not a wrong ruling. The cost is
$\lambda \cdot \mathrm{seats} \cdot \mathrm{attempts}$ in slashes plus
the lost fees, so silence is priced and bounded.

\subsubsection{Frivolous filing}
Every filing and every appeal pays a full tender up front
\eqref{eq:tender}. A spam attack pays $N_i f$ per round it forces the
pool to adjudicate, and drawn jurors keep the fees regardless of the
ruling.

\section{Token model and economics}
\label{sec:econ}

\subsection{Two mints}
Each Subaccord binds two mints at creation. The staking mint is the
collateral: it sets sortition weight and carries the slash exposure.
The fee mint is the compensation: it carries juror fees, appeal bonds,
and bounties. USDC is the common default. Both mints are arbitrary SPL
tokens, and there is no Accord protocol token in version 1. This split
follows one rule: stake is what a juror can lose, fees are what a juror
can earn, and an application wants to tune the two independently. A
pool of dentists can demand a professional-society token as collateral
and pay USDC. A token registry can take its own token as collateral at
a high $S_{\min}$.

\subsection{Fund flows}
All value sits in the two pool vaults. Fees, bonds, and bounties move
only at defined events: filing, appeal, settlement, refund claims, and
fee withdrawal. No fee moves before a dispute is terminal. The
following equality holds at every point in the lifecycle:
\begin{equation}
\begin{split}
  V_{\mathrm{fee}} \;=\; & \sum_{D} \mathrm{fee\_paid}(D)
  + \sum_{j} \mathrm{fees\_earned}(j) \\
  & + \sum_{b} \mathrm{amount}(b)
  + \sum_{D} \Lambda(D),
\end{split}
  \label{eq:invariant}
\end{equation}
that is, the fee vault balance equals the sum of open round pots,
earned but unwithdrawn fees, live bonds, and bounty pools. The stake
vault satisfies the analogous equality against the staked positions.
When both mints are the same token, the protocol guards that the slash
side dominates the fee side of any settlement.

\subsection{Parameter selection}
A pool creator prices three quantities against the dispute values it
expects:
\begin{itemize}
  \item the fee $f$ covers juror attention. The filer pays $(J{+}1)f$
    up front, so $f$ bounds the minimum dispute value worth filing.
  \item the slash $\lambda = \alpha S_{\min}$ must exceed the fee
    share a juror can win, $\lambda > f$. A common range is
    $2 f \le \lambda \le 10 f$.
  \item the ceiling $C^{*}$ of \eqref{eq:ceiling} must exceed the
    largest value the pool will adjudicate, with margin for the
    concentration risk visible in the public stake distribution.
\end{itemize}
Windows trade speed against juror availability. The review, commit,
and reveal windows are pool parameters. The appeal window defaults to
72\,h. Table~\ref{tab:params} collects the defaults and bounds.

\begin{table}[!t]
\caption{Parameters: defaults and bounds.}
\label{tab:params}
\centering
\small
\begin{tabular}{@{}p{24mm}p{26mm}p{20mm}@{}}
\toprule
\textcolor{ink}{\ttfamily\scriptsize PARAMETER} &
\textcolor{ink}{\ttfamily\scriptsize MEANING} &
\textcolor{ink}{\ttfamily\scriptsize DEFAULT / BOUND} \\
\midrule
$J$          & round-0 panel & 3, odd \\
$N_i$        & panel ladder  & $3/7/15/31$ \\
$m$          & max appeals   & 3 \\
$S_{\min}$   & min stake     & $>0$, pool choice \\
$\alpha$     & slash fraction& $\le 1.0$ \\
$\lambda$    & slash amount  & $\alpha S_{\min}$ \\
$f$          & fee per juror & pool choice \\
$\theta$     & reveal quorum & $\tfrac{2}{3}$ \\
$\varepsilon$& median band   & $0.01$ (1\,\%) \\
options      & plurality     & 2 to 8 \\
depth        & accumulator   & 20 ($2^{20}$ seats) \\
appeal window& per round     & 72 h \\
timelock     & parameter update & 48 h \\
\bottomrule
\end{tabular}
\end{table}

\section{Governance and upgrade control}
\label{sec:governance}
Governance is deliberately thin. A Subaccord is a set of parameters
plus an authority key. The authority proposes single-parameter updates
through a 48\,h on-chain timelock. Anyone can observe a proposal and
react. A pool can set the authority to none, which freezes the
parameters forever. The pool identity, the two mints, the aggregation
rule, the coherence band, and the credential gate are immutable from
creation. A dispute freezes the full term set at filing, so no
mid-dispute change can touch a live case.

The program itself carries a Squads multisig upgrade authority with a
pause circuit breaker. Pause blocks new filings and new exposure. It
never consumes an appeal window or freezes a live adjudication. After
sufficient external audit, the upgrade authority goes to none and the
program is frozen. There is no on-chain token voting anywhere in the
stack: rule-making is per pool, and the global protocol ships as fixed
code.

\section{Implementation notes}
\label{sec:implementation}
Accord is a single Anchor program on Solana \cite{solana}. State lives in
program-derived accounts seeded by owner and purpose, so every account is
address-derived and program-owned. Two constraints of Solana shaped the design
at its initial implementation. First, the 1{,}232-byte transaction packet
forces one seat per draw transaction and the accumulator's $O(\log n)$ proof
shape. Second, rent makes every account cost its bytes on chain: a juror
position rents for about ~0.0007 SOL, paid by the juror when staking, and
free-list slot recycling keeps the accumulator from growing without bound. All
progress instructions are permissionless cranks, so no operator owns dispute
advancement. A dispute's compute cost concentrates in Merkle verification and
settlement, both linear in panel size, and settlement writes are batched per
round. The juror-set root is reconstructible from public account state by
anyone, which is the audit path for the accumulator. Ratio parameters
($\alpha$, $\theta$, $\varepsilon$) are stored on chain as integer basis
points.

\section{Evaluation}
\label{sec:evaluation}
Three offline checks ground the quantitative claims of
Sections~\ref{sec:accuracy} and~\ref{sec:attack}. The two curve families are
analytic; the sortition check simulates the exact sampler. None of this is an
on-chain benchmark, and none of it substitutes for a cryptographic proof of the
accumulator or the VRF construction; it checks that the stated approximations
match the mechanism as specified.

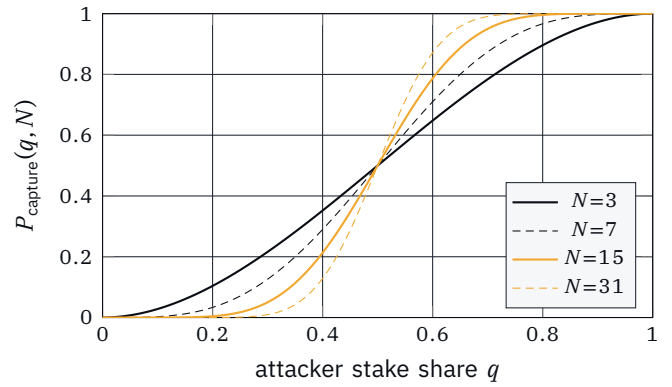
\begin{figure}[!t]
\centering
\begin{tikzpicture}
\begin{axis}[
  width=\columnwidth, height=5.6cm,
  xlabel={attacker stake share $q$},
  ylabel={$P_{\mathrm{capture}}(q,N)$},
  xmin=0, xmax=1, ymin=0, ymax=1,
  grid=major, grid style={draw=hairline},
  tick label style={font=\sffamily\scriptsize},
  label style={font=\sffamily\small},
  legend style={font=\sffamily\scriptsize, draw=hairline,
                fill=paperlight, inner sep=2pt},
  legend pos=south east]
\addplot[ink, thick]           table [x=q, y=n3]  {evaluation/capture.dat};
\addplot[ink, densely dashed]  table [x=q, y=n7]  {evaluation/capture.dat};
\addplot[amber, thick]         table [x=q, y=n15] {evaluation/capture.dat};
\addplot[amber, densely dashed] table [x=q, y=n31] {evaluation/capture.dat};
\legend{$N{=}3$,$N{=}7$,$N{=}15$,$N{=}31$}
\end{axis}
\end{tikzpicture}
\caption{Probability that an attacker controls a strict majority of a
panel, from the independent-draw approximation of \eqref{eq:capture}.
Below $q = 1/2$ larger panels suppress capture; above it they harden
it. The exact distinct-juror sampler deviates from these curves only
in the direction and magnitude bounded in
Figure~\ref{fig:sortition}.}
\label{fig:capture}
\end{figure}

Figure~\ref{fig:capture} plots \eqref{eq:capture}. The reading is direct: at $q
= 0.4$ a 31-seat panel falls to an attacker majority in about 13 percent of
draws, against 35 percent for a 3-seat panel; at $q = 0.6$ the relation inverts
and the large panel favors the attacker (87 against 65 percent). The
security-value ceiling of \eqref{eq:ceiling} is a statement about $q$: a pool
whose stake concentration puts a plausible $q$ above one half has no ceiling
worth pricing, which is why stake distribution is public and priced before
filing.

\subsection{Panel accuracy}
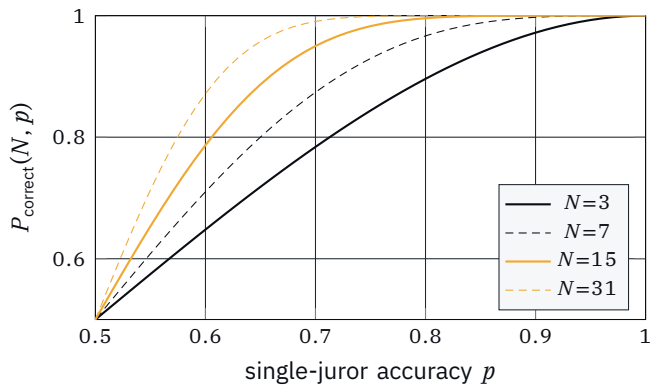
\begin{figure}[!t]
\centering
\begin{tikzpicture}
\begin{axis}[
  width=\columnwidth, height=5.6cm,
  xlabel={single-juror accuracy $p$},
  ylabel={$P_{\mathrm{correct}}(N,p)$},
  xmin=0.5, xmax=1, ymin=0.5, ymax=1,
  grid=major, grid style={draw=hairline},
  tick label style={font=\sffamily\scriptsize},
  label style={font=\sffamily\small},
  legend style={font=\sffamily\scriptsize, draw=hairline,
                fill=paperlight, inner sep=2pt},
  legend pos=south east]
\addplot[ink, thick]           table [x=p, y=n3]  {evaluation/accuracy.dat};
\addplot[ink, densely dashed]  table [x=p, y=n7]  {evaluation/accuracy.dat};
\addplot[amber, thick]         table [x=p, y=n15] {evaluation/accuracy.dat};
\addplot[amber, densely dashed] table [x=p, y=n31] {evaluation/accuracy.dat};
\legend{$N{=}3$,$N{=}7$,$N{=}15$,$N{=}31$}
\end{axis}
\end{tikzpicture}
\caption{Probability of a correct panel majority,
\eqref{eq:condorcet}, under independent honest errors. Every curve
starts at $1/2$ for $p = 1/2$: a pool of coin-flipping jurors has no
aggregate advantage at any panel size.}
\label{fig:accuracy}
\end{figure}

Figure~\ref{fig:accuracy} plots \eqref{eq:condorcet} over the appeal ladder. At
$p = 0.55$ the panel majority is correct 57 percent of the time at $N{=}3$ and
71 percent at $N{=}31$; at $p = 0.9$ the round-0 panel is already at 97 percent
and the ladder adds little. The figure is the paper's contract made visible:
the mechanism converts juror quality into correctness, it does not create
either.

\subsection{Sortition distribution}
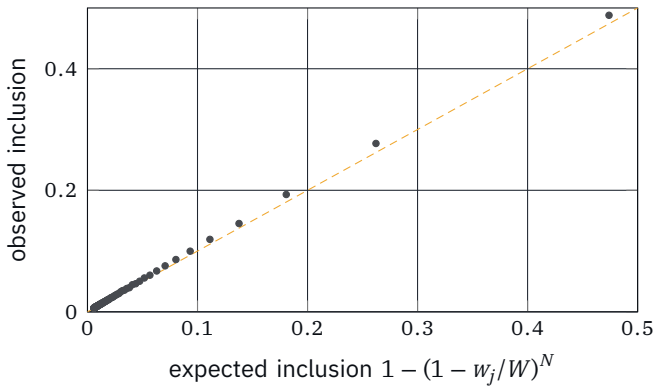
\begin{figure}[!t]
\centering
\begin{tikzpicture}
\begin{axis}[
  width=\columnwidth, height=5.6cm,
  xlabel={expected inclusion $1-(1-w_j/W)^{N}$},
  ylabel={observed inclusion},
  xmin=0, xmax=0.5, ymin=0, ymax=0.5,
  grid=major, grid style={draw=hairline},
  tick label style={font=\sffamily\scriptsize},
  label style={font=\sffamily\small}]
\addplot[only marks, mark=*, mark size=1.2pt, ink!75]
  table [x=expected, y=observed] {evaluation/sortition.dat};
\addplot[amber, densely dashed, domain=0:0.5] {x};
\end{axis}
\end{tikzpicture}
\caption{Panel-inclusion frequency of the exact distinct-juror sampler
against the independent-draw expectation, 100-juror Zipf pool,
$N = 3$, 200{,}000 simulated panels. The dashed line is $y = x$. The
upward residual for heavy jurors is the without-replacement effect of
the collision retry.}
\label{fig:sortition}
\end{figure}

Figure~\ref{fig:sortition} compares the exact sampler---interval
landing with collision retry, as in \eqref{eq:seed} and
\eqref{eq:interval}---against the independent-draw formula of
\eqref{eq:sortition}. The test pool holds 100 jurors with Zipf weights
($w_i \propto 1/i$), the panel size is 3, and the simulation runs
200{,}000 draws. Observed inclusion tracks the expected curve; the
systematic residual runs upward for heavy jurors (the heaviest seat
lands 2.9 percent more often than the independent formula predicts,
the largest deviation across the pool is 14 percent) because a retried
seat redraws among unseated jurors only. This is the deviation the
approximation language in Section~\ref{sec:draw} admits, measured.

\section{Future work}
\label{sec:future}
\begin{itemize}
  \item a validity proof (SNARK) of accumulator root correctness, which
    removes the indexer liveness assumption;
  \item an encrypted vote tally through Arcium, which removes part of
    the bribery and coordination surface;
  \item an anti-pre-revelation bounty after Buterin \cite{antipre}:
    any observer may report a juror's disclosed vote and salt to the
    round, and if the on-chain reveal matches, the pre-revealing juror
    is slashed and the reporter earns a bounty---which turns every
    verifiable receipt into evidence against its issuer and pushes
    vote buying back onto unverifiable promises;
  \item threshold re-encryption or TEE evidence delivery, which removes
    the trusted evidence operator;
  \item a staking-tranche model, which raises the price of
    the pseudonymous key admission;
  \item a round-scaled appeal bounty, so deep appeals stay funded as
    the tender grows exponentially;
  \item further aggregation rules, for example instant-runoff over
    ranked votes;
  \item an epoch machinery that anchors stake for a draw without a
    freeze;
  \item protocol fees paid from earnings of jurors.
  \item a protocol token, only if a function for it survives review.
    None is assumed in this paper.
\end{itemize}

\section{Applications}
\label{sec:apps}
Table~\ref{tab:apps} lists the four shapes that consume the primitive
today or carry a written specification. Two exist as programs with
their own specifications along the accord program in the same repository.

Canon is a curated-list factory where each challenge files a keep-or-remove
dispute and deposits move with the ruling.

Synod is an N-party escrow where the parties themselves are the options and the
prevailing party takes the pot.

Beyond these, the same integration pattern covers freelancing escrow, insurance
claims, prediction-market resolution, grant accountability, and content
moderation. Each application owns its deposits and its enforcement. Accord only
delivers a verdict.

\begin{table}[!ht]
\caption{Reference applications.}
\label{tab:apps}
\centering
\small
\begin{tabular}{@{}p{21mm}p{24mm}p{26mm}@{}}
\toprule
\textcolor{ink}{\ttfamily\scriptsize ARBITRABLE} &
\textcolor{ink}{\ttfamily\scriptsize QUESTION} &
\textcolor{ink}{\ttfamily\scriptsize PROGRAM OWNS} \\
\midrule
Canon (built)   & keep or remove a list item & item lifecycle, deposits \\
Synod (built)   & which party prevails & N-party pot, roster \\
Verdict gate    & should this action run & sole execution key \\
Expert pool     & credential admission & the Subaccord itself \\
\bottomrule
\end{tabular}
\end{table}

\section{Limitations}
\label{sec:limitations}
Security and correctness are separate claims. Security---that the
recorded ruling is the honest aggregation of committed votes under
authenticated randomness---is conditional on an honest majority of
staked collateral and on the cryptographic assumptions of
Section~\ref{sec:threat}. Correctness---that the ruling identifies the
true answer---additionally requires juror information quality above
random, $p > 1/2$, and enough error independence for
\eqref{eq:condorcet} to mean anything; correlated misinformation
shrinks the effective panel (Section~\ref{sec:accuracy}). A stake
majority captures its pool, and appeals do not repair it; Accord
prices capture through per-pool collateral instead of a fork defense.
Juror admission is pseudonymous: distinct keys are not distinct
humans, and a credential gate raises competence but not independence.
Randomness and commitment integrity are cryptographic; randomness
availability, evidence confidentiality, and evidence completeness
depend on external operators, and those failures end in a refund
except where the information layer itself is subverted. Commit--reveal
limits observation and vote copying, not bribery: a juror can prove
its committed vote by disclosing its vote and salt, and after reveal
every vote is public and attributable. Finally, Accord
emits rulings and enforces nothing: an Arbitrable that ignores its own
read-back degrades the oracle to an advisory signal.

\section{Conclusion}
Accord packages Schelling-point arbitration as a Solana primitive: a
stake-weighted, distinct jury drawn by an authenticated VRF over a
canonical accumulator, commit--reveal votes with two aggregation rules,
flat incoherence slashing settled at finality, an exponential appeal
ladder with bonds and flip bounties, and a two-mint per-pool economy
that separates what a juror can lose from what a juror can earn. The
integration surface is two CPI instructions, every residual trust
assumption is stated with a price, and the paper separates what the
mechanism guarantees---authenticated aggregation---from what it can
only condition on---juror information quality. The mechanism inherits
six years of
Kleros production history and adapts it to the constraints of Solana.
Disputes that a smart contract cannot judge now have a native place to
go.

\end{document}